# Voltage-Induced Ferromagnetic Resonance in Magnetic Tunnel Junctions

Jian Zhu,[1] J. A. Katine,[2] Graham E. Rowlands,[1] Yu-Jin Chen,[1] Zheng Duan,[1] Juan G. Alzate,[3] Pramey Upadhyaya,[3] Juergen Langer,[4] Pedram Khalili Amiri,[3] Kang L. Wang,[3] and Ilya N. Krivorotov[1]

[1]*Department of Physics and Astronomy, University of California, Irvine, California 92697, USA*
[2]*Hitachi Global Storage Technologies, 3403 Yerba Buena Road, San Jose, California 95135, USA*
[3]*Department of Electrical Engineering, University of California, Los Angeles, California 90095, USA*
[4]*Singulus Technologies, 63796 Kahl am Main, Germany*

We demonstrate excitation of ferromagnetic resonance in CoFeB/MgO/CoFeB magnetic tunnel junctions (MTJs) by the combined action of voltage-controlled magnetic anisotropy (VCMA) and spin transfer torque (ST). Our measurements reveal that GHz-frequency VCMA torque and ST in low-resistance MTJs have similar magnitudes, and thus that both torques are equally important for understanding high-frequency voltage-driven magnetization dynamics in MTJs. As an example, we show that VCMA can increase the sensitivity of an MTJ-based microwave signal detector to the sensitivity level of semiconductor Schottky diodes.

Excitation of sub-nanosecond magnetic dynamics by electric field is a grand challenge in the field of spintronics. The ability to perform high-speed manipulation of magnetization by electric fields rather than by current-induced spin torques or magnetic fields would greatly reduce ohmic losses and thereby improve the performance of spintronic devices such as non-volatile magnetic memory. In this Letter we demonstrate excitation of GHz-range magnetization dynamics in nanoscale MTJs by the combined action of ST arising from spin-polarized current [1–9] and VCMA induced by electric field [10–20]. We show that ST and VCMA torques have similar magnitudes in MTJs with ultra-thin ($< 1$ nm) tunnel barriers for both DC and microwave-frequency voltages. Our results demonstrate that a description of voltage-induced high-speed magnetic dynamics in MTJs should generally include not only ST but also VCMA terms. As an example, we show that VCMA can increase the sensitivity of a ST microwave signal detector, raising it to the sensitivity level of semiconductor Schottky diodes.

We make measurements of voltage-driven magnetization dynamics in 150×70 nm² elliptical MTJ nanopillars shown schematically in Fig. 1. The nanopillars are patterned by ion milling from (bottom lead)/ Ta(5)/ PtMn(15)/ $Co_{70}Fe_{30}$(2.3)/ Ru(0.85)/ $Co_{40}Fe_{40}B_{20}$(2.4)/ MgO(0.83)/ $Co_{20}Fe_{60}B_{20}$(1.58)/ Ta(5)/ (cap) multilayer (thicknesses in nm) with resistance-area product of 3.5 $\Omega \cdot \mu m^2$ deposited by magnetron sputtering in a Singulus TIMARIS system. Prior to patterning, the multilayers are annealed for 2 hours at 300 °C in a 1 Tesla in-plane magnetic field that sets the pinned layer exchange bias direction parallel to the long axis of the nanopillars (the $\hat{x}$-axis of our Cartesian coordinate system). The $Co_{20}Fe_{60}B_{20}$ free layer exhibits perpendicular magnetic anisotropy [21], and the equilibrium direction of its magnetization at zero field is normal to the sample plane ($\hat{z}$-axis). Measurements of the nanopillar conductance $G$ versus $\hat{x}$-axis magnetic field $H_x$ (defined as the sum of the external $\hat{x}$-axis field and the stray field from the polarizer) confirm the out-of-plane orientation of the free layer magnetization at $H_x = 0$ (see Fig. 1). The MTJ conductance exhibits cosine dependence on the angle $\phi$ between the magnetic moments of the free and pinned layers [22]:

$$G = G_0 \left(1 + P^2 \cos(\phi)\right), \quad (1)$$

where $P$ is the current spin polarization. It follows that the shape of the $G(H_x)$ curve is identical to that of the $M_x(H_x) = M_s \cos(\phi)$ hysteresis loop, where $M_s$ is saturation magnetization of the free layer and $M_x$ is projection of magnetization onto the $\hat{x}$ axis. The $G(H_x)$ curves in Fig. 1 are hard-axis hysteresis loops, the slopes of which depend on the applied direct voltage bias $V_{dc}$ by virtue of VCMA [17–20]. As discussed in the supplementary online material [23], ST from $V_{dc}$ does not significantly affect the $G(H_x)$ hysteresis loop.

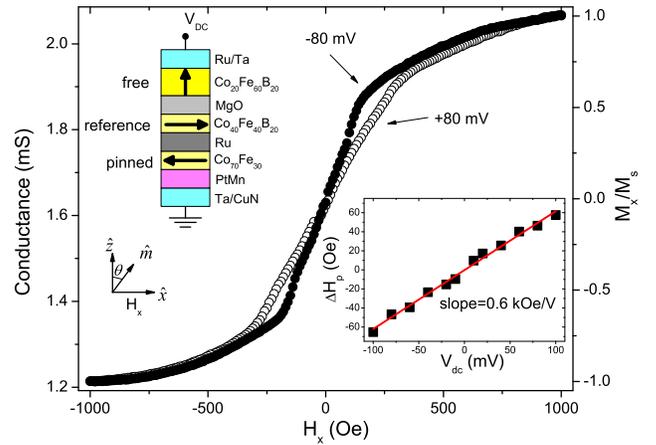

FIG. 1. (color online). Hysteresis loop of MTJ conductance $G$ versus in-plane magnetic field $H_x$ measured under ± 80 mV direct voltage bias $V_{dc}$. The left inset shows schematic of the MTJ nanopillar. The right inset displays variation of the effective perpendicular anisotropy field $\Delta H_p(V_{dc})$ with DC voltage extracted from the $G(H_x)$ hysteresis loops via Eq.(2) (squares) and a linear fit to the data (line).

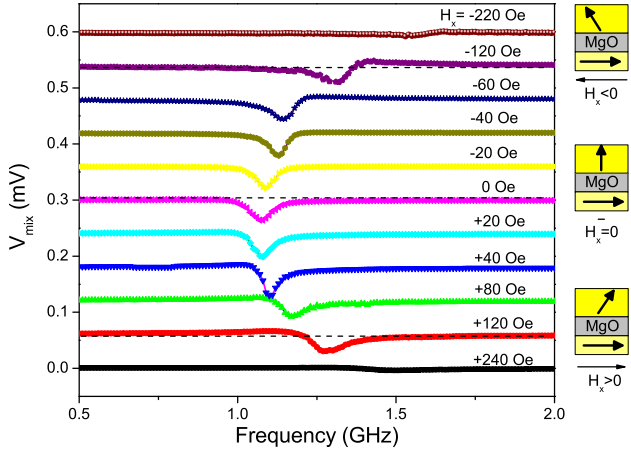

FIG. 2. (color online). Rectified voltage $V_{mix}$ versus frequency $f$ of the applied microwave signal measured at several values of the in-plane magnetic field $H_x$. The FMR response curves $V_{mix}(f)$ at different fields are vertically offset by 0.06 mV. At $H_x=0$, $V_{mix}(f)$ is a symmetric Lorentzian curve, while at $|H_x|>0$, $V_{mix}(f)$ is a sum of symmetric and antisymmetric Lorentzians. The antisymmetric part of $V_{mix}(f)$ changes sign upon reversal of $H_x$, while the symmetric part is negative for any $H_x$.

The effective perpendicular magnetic anisotropy energy per unit volume of the free layer, $E_p$, includes magnetocrystalline, surface, and magnetostatic anisotropy contributions and can be calculated from the area under the hysteresis loop $M_x(H_x)$ in Fig. 1:

$$E_p = \int_0^{M_s} H_x(M_x)dM_x = \frac{1}{2}M_s H_p, \quad (2)$$

where $H_p$ is the effective perpendicular anisotropy field. The inset in Fig. 1 displays the variation of the perpendicular anisotropy field with DC voltage $\Delta H_p(V_{dc}) = H_p(V_{dc}) - H_p(0)$. This dependence is well fit by a straight line with the slope of 0.6 kOe/V, which corresponds to magnetic anisotropy energy per area per electric field of 37 fJ/(V·m). This value is similar to the magnitude of VCMA reported by other groups [15, 17].

We study the effect of high-frequency voltage oscillations on magnetization of the MTJ free layer using a ferromagnetic resonance (FMR) technique. In this technique, magnetization oscillations are excited by a microwave voltage applied to the MTJ, and the amplitude of magnetization precession is detected electrically via the rectified voltage $V_{mix}$ generated by the MTJ [24, 25]. This technique is commonly called spin torque FMR (ST-FMR), since the magnetization precession of the free layer is driven by ST from the microwave current $I_{ac}\cos(2\pi ft)$ flowing through the MTJ [26]. However, as we show below, magnetization oscillations in our MTJ samples are excited not only by ST from $I_{ac}$ but also by oscillating VCMA induced by the applied microwave voltage $V_{ac}$. In this method, the rectified voltage $V_{mix} = \frac{1}{2}I_{ac}R_{ac}\cos(\psi)$ arises from mixing of the microwave current $I_{ac}\cos(2\pi ft)$ and tunneling magnetoresistance oscillations $R_{ac}\cos(2\pi ft + \psi)$ resulting from the free layer's magnetization precession [25]. We sweep the microwave drive frequency $f$ and measure ST-FMR response curve $V_{mix}(f)$. In general, peaks in the $V_{mix}(f)$ response curve arise from resonant excitation of spin wave eigenmodes of the MTJ, and analysis of the line shape of these ST-FMR resonances gives the value of Gilbert damping and the magnitude of voltage-induced torques (including ST) acting on magnetization [26].

The measured dependence of $V_{mix}(f)$ is shown in Fig. 2 for several values of $H_x$. The negative peaks in $V_{mix}(f)$ arise from excitation of the quasi-uniform FMR mode of the free layer [25]. For all measurements, the applied microwave power is −36 dBm, which is small enough to excite the free layer precession in the linear regime. At $H_x = 0$, the magnetization of the free layer is perpendicular to the sample plane ($\theta = 0$), and the ST-FMR line shape $V_{mix}(f)$ is well described by a symmetric Lorentzian. For non-zero $H_x$, the magnetization of the free layer is tilted away from the sample normal, and $V_{mix}(f)$ develops asymmetry. In this regime, the ST-FMR line shape is well fit by a sum of symmetric and antisymmetric Lorentzians with identical resonance frequencies $f_0$:

$$V_{mix}(f) = \frac{V_s}{1 + (f - f_0)^2/\sigma^2} + \frac{V_a(f - f_0)/\sigma}{1 + (f - f_0)^2/\sigma^2}, \quad (3)$$

where $V_s$ and $V_a$ are the symmetric and antisymmetric term amplitudes, and $\sigma$ is the half width at half maximum. As is evident from Fig. 2 (compare ST-FMR curves at $H_x = \pm 120$ Oe), the antisymmetric part of the ST-FMR line shape, $V_a$, changes sign upon reversal of $H_x$.

The symmetric component of the ST-FMR curve arises from in-plane ST $\vec{\tau}_i \sim \hat{\mathbf{m}} \times (\hat{\mathbf{e}}_p \times \hat{\mathbf{m}})$ that lies in the plane defined by the magnetic moments of the free and the pinned layers, where $\hat{\mathbf{m}}$ is the unit vector in the direction of the free layer's magnetization and $\hat{\mathbf{e}}_p$ is the unit vector in the direction of the polarizer's magnetization. An antisymmetric component of the ST-FMR curve has been previously observed in nanoscale MTJs at non-zero DC bias voltages, and was shown to arise from a field-like ST, $\vec{\tau}_f \sim \hat{\mathbf{m}} \times \hat{\mathbf{e}}_p$ [27, 28]. The antisymmetric component was observed to vanish at zero bias voltage, which was explained by quadratic dependence of $\vec{\tau}_f$ on the voltage bias [27]. Hence, our observation of a non-zero antisymmetric component of the ST-FMR response curve at zero DC voltage is surprising.

In order to understand the origin of the ST-FMR line shape asymmetry in our MTJ system, we numerically simulate [23] the resonance curve by solving the Landau-Lifshitz-Gilbert equation of motion of the free layer in the macrospin approximation [29]:

$$\frac{d\hat{\mathbf{m}}}{dt} = -\gamma \hat{\mathbf{m}} \times \vec{H}_\text{eff} + \alpha \hat{\mathbf{m}} \times \frac{d\hat{\mathbf{m}}}{dt} - \gamma \frac{J\hbar}{2e\,M_s\,d}$$
$$\frac{P}{1 + P^2 \cos(\phi)} [\hat{\mathbf{m}} \times (\hat{\mathbf{m}} \times \hat{\mathbf{e}}_p) + b_f \hat{\mathbf{m}} \times \hat{\mathbf{e}}_p]. \quad (4)$$

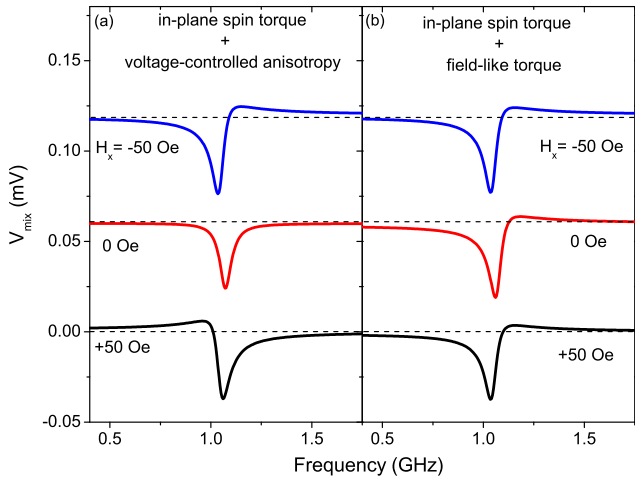

FIG. 3. (color online). Macrospin simulations of ST-FMR line shape assuming non-zero in-plane ST and (a) non-zero VCMA ($\Delta H_\perp = 4$ Oe) and zero field-like ST ($b_f = 0$), (b) zero VCMA ($\Delta H_\perp = 0$ Oe) and non-zero field-like ST ($b_f = 0.5$). The curves for different values of $H_x$ are vertically offset by 0.06 mV.

In Eq.(4), $\vec{H}_\text{eff}$ is the effective field including contributions from the external, magnetostatic, and perpendicular magnetic anisotropy fields, $\gamma$ is the gyromagnetic ratio, $\alpha$ is the Gilbert damping parameter, $J\hbar/2e$ is the microwave spin current density, $d$ is the thickness of the free layer, and $b_f$ is the ratio of the magnitudes of the in-plane ST and field-like ST.

For the approximately quadratic dependence of $\vec{\tau}_f$ on the current bias observed in previous experiments [27, 28, 30], $b_f$ is proportional to $J$ and thus $\vec{\tau}_f$ can be neglected at zero current bias. However, we include $\vec{\tau}_f$ in the simulations since a linear dependence of $\vec{\tau}_f$ on current (corresponding to $b_f = const$) is, in principle, possible for MTJs with different free and pinned ferromagnetic layers [31]. Such a linear dependence of $\vec{\tau}_f$ on current would give rise to an antisymmetric component of the ST-FMR response curve at zero bias.

In our simulations, both the charge current density $J = J_0 \sin(2\pi f t)$ and the effective magnetic field $\vec{H}_\text{eff}$ are generally time-dependent. The time-dependence of the effective magnetic field arises from the VCMA, giving the time-dependent perpendicular anisotropy field $H_{\perp z} = (H_{\perp 0} + \Delta H_\perp \sin(2\pi f t)) \cos(\theta)$, where $H_{\perp 0}$ is the anisotropy field at zero bias and $\Delta H_\perp$ is the VCMA field. The simulated ST-FMR response curves shown in Fig. 3(a) are calculated assuming non-zero in-plane ST and non-zero VCMA ($\Delta H_\perp = 4$ Oe), but zero field-like ST ($b_f = 0$). In these simulations, we use $H_{\perp 0} = 11,362$ Oe, $\alpha = 0.030$, and $P = 0.54$ as determined by fitting the resonance frequency, the line width and the amplitude of the ST-FMR resonance curve measured at $H_x = 0$. Allowing for uncertainty in the sample volume, we note that $P = 0.54$ agrees well with the value $P = 0.52$ calculated from the MTJ conductance using Eq.(1). In the simulations, we use demagnetization factors $N_x = 0.014, N_y = 0.040, N_z = 0.946$ obtained from the assumption that the free layer is an elliptic cylinder [32]. The saturation magnetization $M_s$ is taken to be 950 emu/cm$^3$ as measured by vibrating sample magnetometry, and $\Delta H_\perp = 4$ Oe is calculated from the slope of $\Delta H_p(V_{dc})$ line in the inset of Fig. 1. Fig. 3(a) shows that the simulated ST-FMR line shapes reproduce the symmetry of the measured curves: (i) the line shape is symmetric at $H_x = 0$ and (ii) the line shape is asymmetric with the antisymmetric part changing sign upon reversal of $H_x$ for $|H_x| > 0$. In contrast, simulations with a non-zero filed-like torque $\vec{\tau}_f$ do not reproduce the symmetry of the measured line shapes. Fig. 3(b) shows the simulation results for non-zero in-plane ST and non-zero field-like ST ($b_f = 0.5$) but zero VCMA ($\Delta H_\perp = 0$). The simulated ST-FMR line shapes are asymmetric for all applied field values, and the sign of the antisymmetric part does not depend on the applied field direction. Therefore, our simulations demonstrate that the observed ST-FMR line shape asymmetry arises from VCMA, not field-like ST.

The antisymmetric part of the ST-FMR line shape calculated with the VCMA value derived from DC measurements ($\Delta H_\perp = 4$ Oe) has similar magnitude to that of the measured ST-FMR response curve. This implies that the magnitude of VCMA is comparable for microwave and DC voltages. Since the measured magnitudes of the symmetric and antisymmetric parts of the ST-FMR line shapes are similar for $\theta \neq 0$, we conclude that ST and VCMA torque generally have similar magnitudes in our MTJs.

Although the ST-FMR line shape is well reproduced by the simulations with the VCMA drive term, the dependence of the simulated resonance frequency $f_0$ on $H_x$ does not qualitatively agree with the measurements. The macrospin simulations predict decreasing $f_0$ with increasing $|H_x|$ as shown in Fig. 4 (a,c), while the measured $f_0(|H_x|)$ shows the opposite trend (Fig. 2). We find that this discrepancy arises from the simplified form of the perpendicular anisotropy energy used in the simulations. The thickness of the free layer in our samples is intentionally tuned to a value at which the uniaxial perpendicular anisotropy and the easy-plane shape anisotropy have comparable magnitudes but opposite signs. Therefore, the leading terms proportional to $\sin^2(\theta)$ in the angular dependence of both anisotropy energy expansions nearly cancel each other and the higher order term proportional to $\sin^4(\theta)$ becomes important [33].

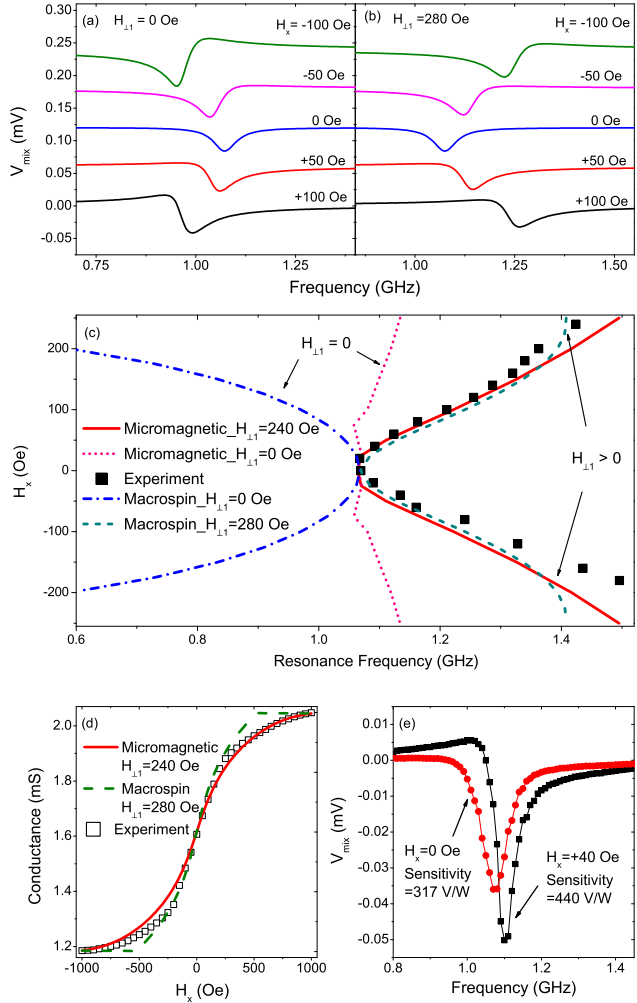

FIG. 4. (color online). Macrospin simulations of ST-FMR response curves with (a) zero ($H_{\perp 1} = 0$) and (b) positive ($H_{\perp 1} = 280$ Oe) second-order perpendicular anisotropy. The curves are vertically offset by 0.06 mV. (c) Macrospin and micromagnetic simulations of ST-FMR resonance frequency versus $H_x$ for zero and positive $H_{\perp 1}$ (lines) shown together with the experimental data (solid squares). (d) Macrospin and micromagnetic simulations (lines) of the $G(H_x)$ hysteresis loop using the magnetic anisotropy parameters given by the best fits in (c). Squares are experimental data measured at zero bias voltage. (e) Sensitivity of the MTJ to a microwave signal is enhanced by VCMA when the free layer magnetization is tilted from the sample normal by magnetic field $H_x = +40$ Oe.

In order to study the impact of the second-order perpendicular anisotropy term on $f_0(H_x)$, we repeat our macrospin simulations with non-zero first- and second-order perpendicular anisotropy terms: $E_\perp = K_1 \sin^2(\theta) + K_2 \sin^4(\theta)$, which gives $H_{\perp z} = [H_{\perp 0} + H_{\perp 1}(1 - \cos^2(\theta))] \cos(\theta)$ at zero bias, where $H_{\perp 0} = 2K_1/M_s$ and $H_{\perp 1} = 4K_2/M_s$. Our simulations show that adding a small positive second-order anisotropy field $H_{\perp 1}$ changes the shift of the resonance frequency with $|H_x|$ from negative to positive, and the results from simulations at $H_{\perp 1} = 280$ Oe are in good quantitative agreement with the measured $f_0(H_x)$ (Fig. 4 (b,c)).

We also make micromagnetic simulations of $f_0(H_x)$ using MicroMagus simulator and find the micromagnetic $f_0(H_x)$ to be in qualitative agreement with the macrospin simulation results. Fig. 4 (c) shows that micromagnetic simulations performed with $H_{\perp 1} = 0$ give $f_0(H_x)$ inconsistent with the experimental data, while those performed with $H_{\perp 0} = 11,515$ Oe and $H_{\perp 1} = 240$ Oe reproduce the measured $f_0(H_x)$ well. Micromagnetic and macrospin simulations of the $G(H_x)$ hysteresis loop made with the values of $H_{\perp 0}$ and $H_{\perp 1}$ obtained from best fits of $f_0(H_x)$ are also in good agreement with the experimental data as shown in Fig. 4(d).

The measured amplitudes of the symmetric and antisymmetric parts of the ST-FMR line shape are similar to each other for $|H_x| > 0$. This implies that the magnitude of VCMA torque is similar to that of ST, and thus both torques are important for the description of voltage-driven magnetization dynamics in MTJs, such as ST-induced magnetization reversal [34, 35]. As an example, Fig. 4 (e) shows that VCMA can significantly increase the sensitivity of an MTJ-based microwave signal detector [24]. Indeed, the maximum rectified voltage generated by the MTJ in response to applied microwave signal increases by 39% when $H_x$ increases from 0 Oe to +40 Oe [23]. This increase in the detector sensitivity is mainly due to the antisymmetric part of the ST-FMR line shape, which is induced by VCMA. With the VCMA contribution, the MTJ microwave detector sensitivity reaches 440 V/W — a value similar to that of semiconductor Schottky diodes ($\sim 5 \times 10^2$ V/W [36]).

In conclusion, we demonstrate excitation of GHz-range magnetization dynamics in nanoscale MTJs by the combined action of spin torque and voltage-controlled magnetic anisotropy. Our work shows that the magnitudes of high-frequency spin torque and voltage-controlled magnetic anisotropy torque in MgO-based MTJs can be similar to each other, and thus that quantitative descriptions of voltage-driven magnetization dynamics in MTJs should generally include both torque terms. We show that voltage-controlled magnetic anisotropy can increase the sensitivity of spin torque MTJ microwave signal detectors to the sensitivity level of Schottky diode detectors.

This work was supported by DARPA Grant Nos. HR0011-09-C-0114 and HR0011-10-C-0153, by NSF Grant Nos. DMR-0748810 and ECCS-1002358, and by the Nanoelectronics Research Initiative through the

Western Institute of Nanoelectronics.

---

# Voltage-Induced Ferromagnetic Resonance in Magnetic Tunnel Junctions

Jian Zhu[1], Jordan A. Katine[2], Graham E. Rowlands[1], Yu-Jin Chen[1], Zheng Duan[1], Juan Alzate[3], Pramey Upadhyaya[3], Juergen Langer[4], Pedram Khalili Amiri[3], Kang L. Wang[3], and Ilya N. Krivorotov[1]

1 Department of Physics and Astronomy, University of California, Irvine, California 92697, USA

2 Hitachi Global Storage Technologies, 3403 Yerba Buena Road, San Jose, California 95135, USA

3 Department of Electrical Engineering, University of California, Los Angeles, California 90095, USA

4 Singulus Technologies, 63796 Kahl am Main, Germany

## Supplementary Material

**1. Macrospin Simulations**

We simulate dynamics of the MTJ free layer by numerically solving Landau-Lifshitz-Gilbert (LLG) equation with the spin torque term in the macrospin approximation:

$$\frac{d\hat{m}}{dt} = -\gamma \hat{m} \times \vec{H}_{eff} + \alpha \hat{m} \times \frac{d\hat{m}}{dt} - \gamma \frac{J\hbar}{2eM_s d} \frac{P}{1+P^2 \cos(\phi)} [\hat{m} \times (\hat{m} \times \hat{e}_p) + b_f \hat{m} \times \hat{e}_p], \quad (S1)$$

where $\hat{m}(t)$ is the unit vector in the direction of the free layer magnetic moment. In the LLG equation, $J(t) = J_0 \cos(2\pi f t)$, where $f$ is the frequency of the applied microwave voltage, $\vec{H}_{eff} = \vec{H}_{ext} + \vec{H}_{demag} + \vec{H}_p$ is the total effective field including the external $\vec{H}_{ext} = H_x \hat{x}$, demagnetizing $\vec{H}_{demag} = -N_x 4\pi M_S m_x \hat{x} - N_y 4\pi M_S m_y \hat{y} - N_z 4\pi M_S m_z \hat{z}$ and perpendicular magnetic anisotropy $\vec{H}_p = (H_\perp + \Delta H_\perp \sin(2\pi f t)) m_z \hat{z}$ fields. Here $H_\perp$ is the static perpendicular anisotropy field at zero voltage bias and $\Delta H_\perp \sin(2\pi f t)$ is the time-dependent perpendicular anisotropy field arising from the applied ac voltage via VCMA. The value of $\Delta H_\perp$ = 4 Oe is derived from the change of slope of the hard axis hysteresis loop with applied dc voltage as shown in Fig. 1. The angular dependence of the perpendicular anisotropy field $H_\perp m_z = [H_{\perp 0} + H_{\perp 1}(1-\cos^2\theta)]\cos\theta$ is derived from the angular dependence of the perpendicular magnetic anisotropy energy density $E_\perp$, in which first- and second-order terms are included: $E_\perp = K_1\sin^2\theta + K_2\sin^4\theta$, where $H_{\perp 0} = 2K_1/M_s$ and $H_{\perp 1} = 4K_2/M_s$. The magnitude of saturation magnetization of the free layer $M_s$ = 950 emu/cm$^3$ is directly measured by vibrating sample magnetometry of unpatterned multilayer annealed under the same conditions as the multilayer used for the nanopillar fabrication. We employ the demagnetization factors $N_x$ = 0.014, $N_y$ = 0.040, $N_z$ = 0.946 appropriate for the 150 × 70 × 1.58 nm$^3$ elliptical cylinder, which are the dimensions of the free layer. The value of the Gilbert damping constant $\alpha$ = 0.030 is determined by fitting the line width of the measured ferromagnetic resonance curve while the degree of spin polarization of the current $P$ = 0.54 is determined from fitting the amplitude of the ferromagnetic resonance curve. The magnitude of the first-order perpendicular anisotropy field $H_{\perp 0}$ = 11,362 Oe is given by the best fit of the resonance frequency at zero field ($H_x$ = 0). The magnitude of the second-order perpendicular anisotropy field $H_{\perp 1}$ = 280 Oe is given by the best fit of the dependence of the resonance frequency on applied magnetic field $f_0(H_x)$.

We numerically solve the LLG equation for a period of time sufficiently long for the free layer magnetic moment $\hat{m}(t)$ to reach the regime of steady-state oscillations. In this steady state regime, we use $\hat{m}(t)$ to calculate the time-dependent MTJ conductance $G = G_0(1 + P^2\cos(\phi)) = G_0(1 + P^2 m_x(t))$, and the time-dependent voltage generated by the MTJ due to the conductance oscillations $V_g(t) = \dfrac{J_0\cos(2\pi f t)}{G_0(1 + P^2 m_x(t))}$.

The mixing voltage generated by the MTJ, $V_{mix}$, at the drive frequency $f$ is calculated as time average of $V_g(t)$ over one period of the magnetization precession. Performing the calculation of $V_{mix}$ as a function of the drive frequency $f$, we obtain the simulated ferromagnetic resonance curve $V_{mix}(f)$ in the macrospin approximation.

**2. Effect of direct voltage on the hysteresis loop of the free layer**

Direct voltage applied to the MTJ generally induces in-plane spin torque, field-like torque and VCMA torque, each of which can potentially alter the hysteresis loop of magnetization versus magnetic field $M_x(H_x)$. However, each type of torque is expected to produce a distinctly different effect on the $M_x(H_x)$ hysteresis loop and thus measurements of $M_x(H_x)$ as a function of voltage can give us information on the type of voltage-induced torques present in the system.

Effective magnetic field arising from the in-plane spin torque is perpendicular to the plane defined by the magnetic moments of the free layer and the polarizer. At zero applied voltage, magnetic moment of the polarizer is parallel to the $x$-axis and the free layer is in the $x$-$z$ plane, where $z$ is the sample normal. Therefore, the effective field form the in-plane spin torque is in the $y$ direction, which gives rise to rotation of the free layer magnetization in the direction perpendicular to the $x$-$z$ plane. As a result, the angle between the magnetic moments of the free layer and the polarizer $\phi$ (and thus $M_x = M_s \cos(\phi)$) remains unchanged to first order in the free layer rotation angle. For this reason, we do not expect the in-plane spin torque to significantly modify the $M_x(H_x)$ hysteresis loop.

The effective field arising from the field-like torque is collinear with the polarizer direction (along the $x$-axis), which is the same as the direction of the applied magnetic field $H_x$. As a result, the field-like torque is expected to shift the hysteresis loop along the magnetic field axis but it cannot change the hard-axis hysteresis loop slope. Since we do not see a significant voltage-induced shift of the measured hysteresis loop along $H_x$, we conclude that field-like torque is relatively small for the range of direct voltages employed in the experiment.

The effective field arising from VCMA is parallel to the $z$-axis and it causes the magnetization of the free layer to rotate in the $x$-$z$ plane thereby changing the angle $\phi$ between the free layer and the polarizer (and thus $M_x = M_s \cos(\phi)$). When we sweep the external magnetic field, the polar angle of the free layer will change and hence change the effective field. Therefore, the shift of the hysteresis loop is not a constant as opposite to the field-like torque. In fact, it will change the slope of the hysteresis loop. That explains that the change of slopes of the hysteresis loop is due to the VCMA.

3. **Field-dependence of the MTJ microwave detector sensitivity**

The sensitivity of a MTJ-based microwave signal detector is defined as $V_{dc}/P_{in}$, where $V_{dc}$ is direct voltage generated by the MTJ and $P_{in}$ is the applied microwave power. In our experiment, the rectified voltage $V_{mix} = (\sqrt{2}/\pi)V_{dc}$ is measured with a lock-in amplifier. Fig.S1 shows the measured dependence of the MTJ sensitivity on in-plane magnetic field $H_x$. The maximum sensitivity is observed at $H_x > 0$ due to a non-zero VCMA torque acting on the free layer magnetization tilted away from the sample normal by $H_x$.

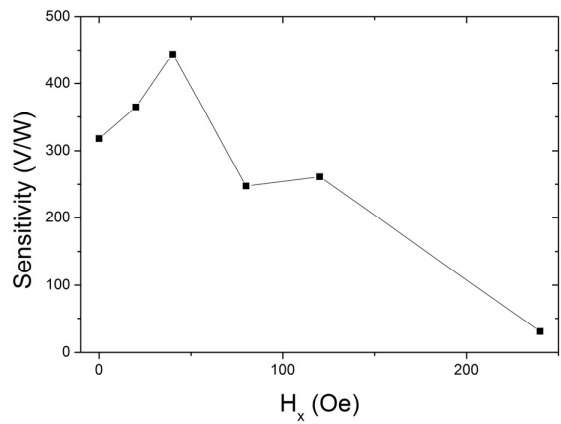

FIG.S1. Sensitivity of the MTJ as a function of in-plane magnetic field.